# Solar Energetic Particle Acceleration at a Spherical Shock with the Shock Normal Angle $\theta_{B_n}$ Evolving in Space and Time

Xiaohang Chen[1], Joe Giacalone[1], and Fan Guo[2]
[1] Lunar and Planetary Laboratory, University of Arizona, Tucson, AZ, USA; xiaohang@lpl.arizona.edu
[2] Los Alamos National Laboratory, Theoretical Division, Los Alamos, NM, USA

## Abstract

We present a 2D kinematic model to study the acceleration of solar energetic particles (SEPs) at a shock driven by a coronal mass ejection. The shock is assumed to be spherical about an origin that is offset from the center of the Sun. This leads to a spatial and temporal evolution of the angle between the magnetic field and the shock-normal direction ($\theta_{Bn}$) as it propagates through the Parker spiral magnetic field from the lower corona to 1 au. We find that the high-energy SEP intensity varies significantly along the shock front due to the evolution of $\theta_{Bn}$. Generally, the west flank of the shock preferentially accelerates particles to high energies compared to the east flank and shock nose. This can be understood in terms of the rate of acceleration, which is higher at the west flank. Double power-law energy spectra are reproduced in our model as a consequence of the local acceleration and transport effects. These results will help us to better understand the evolution of SEP acceleration and provide new insights into large SEP events observed by multiple spacecraft, especially those close to the Sun, such as Parker Solar Probe and Solar Orbiter.

## 1. Introduction

Solar energetic particles (SEPs) are high-energy particles accelerated during eruptive events near the Sun, such as solar flares and coronal mass ejections (CMEs; see reviews by Reames 1999; Desai & Giacalone 2016). Large SEP events are of particular interest to space weather because they pose a serious radiation hazard to astronauts and electronic equipment in space. In the most-intense events, SEPs with sufficiently high energy (about a few GeV nucleon$^{-1}$) can penetrate inside Earth's atmosphere and produce secondary neutrons that can be detected by ground-based neutron monitors. These are known as the ground-level enhancements (GLEs; Forbush 1946; Meyer et al. 1956). It is widely believed that large SEP events, especially GLEs, are associated with collisionless shocks driven by fast and wide CMEs (Kahler et al. 1978; Gosling et al. 1981; Gosling 1993; Lario et al. 2003; Giacalone 2012).

In these events, SEPs are accelerated by a CME-driven shock continuously from the Sun to several astronomical units and spread over a broad range of heliolongitudes. Multispacecraft observations have shown that the time–intensity profiles of a single SEP event often exhibit significant variations at widely separated heliospheric locations that are well organized by magnetic connectivity to the origin of CMEs (Lario et al. 2016). It is not presently clear how the intensity varies along the shock front for any given shock, however. Also, the energy spectra of SEPs usually show an exponential rollover (Ellison & Ramaty 1985) or a double power-law (Band et al. 1993) feature with the break energy depending on the ion charge-to-mass ratio (Cohen et al. 2005; Mewaldt et al. 2005). So far, a number of scenarios are proposed to be important in reproducing double power-law spectra, such as finite acceleration time, finite shock size, shock geometry, adiabatic cooling, and particle escape (e.g., Channok et al. 2005; Li et al. 2009; Schwadron et al. 2015; Zhao et al. 2016; Kong et al. 2019; Fraschetti & Balkanski 2022), which can be classified as either a local acceleration process or a transport effect. However, which one is the dominant factor and how the spectra evolve in space and time as the shock propagates through the interplanetary magnetic fields (IMF) remains unclear.

It is widely believed that diffusive shock acceleration (DSA) is the primary acceleration mechanism in producing energetic particles in many heliophysics and astrophysical systems (Axford et al. 1977; Krymskii 1977; Bell 1978; Blandford & Ostriker 1978; Jokipii 1982). Particles are accelerated as they move across the strong plasma compression at the shock front. For efficient acceleration, the particles must remain near the shock, and this is caused by the scattering within magnetic irregularities in the solar wind. This acceleration process naturally predicts a universal power-law momentum distribution $f \propto p^{-\gamma}$, with the index $\gamma$ only depending on the shock compression ratio (ratio of the downstream to upstream plasma density, which is related to the velocity compression). The acceleration time to a certain energy (acceleration rate) scales inversely with diffusion coefficients that are associated with local magnetic turbulence (e.g., Forman & Drury 1983; Jokipii 1987). The shock geometry also plays a critical role in regulating the acceleration and transport processes (Giacalone et al. 1994; Giacalone 2004; Guo et al. 2010; Kong et al. 2017). Generally, charged particles in the IMF have a much smaller diffusion coefficient perpendicular to the mean magnetic field than that parallel to it. This gives rise to a larger acceleration rate at a quasi-perpendicular shock ($\theta_{B_n} > 45°$; $\theta_{B_n}$ is the angle between the upstream magnetic field and the shock normal) than that at a quasi-parallel shock ($\theta_{B_n} < 45°$). However, the situation is not so simple because, as CME-driven shocks travel in the IMF, the shock-normal angle is neither simply perpendicular nor parallel but evolves in both space and time. For example, for a shock near the Sun, the IMF is almost radial and $\theta_{B_n}$ varies from $\sim 0°$ at the shock nose to $\sim 90°$ near the shock flanks (Figure 1(a)); for the shock at/beyond 1 au, the upstream magnetic field is nearly uniform and $\theta_{B_n} \sim 45°$ (Figure 1(b)). These geometric effects could significantly change the acceleration rate along the shock front and thus account for the longitudinal dependence of SEP observations at 1 au. The duration of magnetic-field line connectivity to the shock also likely plays a significant role since, generally, particles remain relatively close to magnetic lines of force.

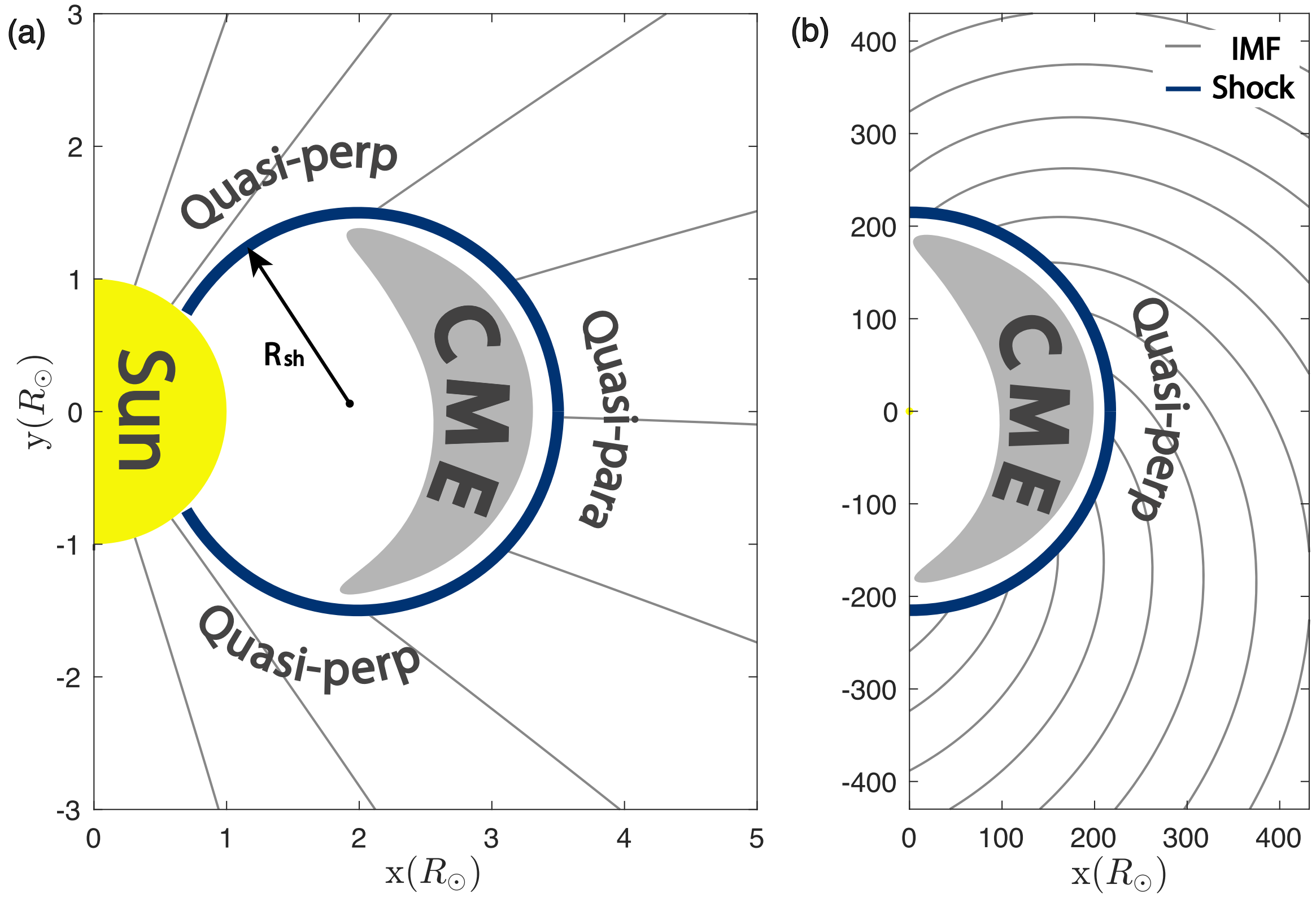

**Figure 1.** IMF topology associated with a CME-driven shock in the regions (a) near the Sun and (b) at 1 au.

Until relatively recently, the problem of particle acceleration at evolving shocks in which $\theta_{Bn}$ varies along the shock front has received little attention. For example, Giacalone (2017) investigated the distribution of energetic particles at a spherical blast wave in a nearly uniform magnetic field where shock-normal angle $\theta_{Bn}$ varied from quasi-parallel at the poles to quasi-perpendicular near the equator. Kong et al. (2017, 2019) studied SEP energization at coronal shocks propagating in the streamer-like magnetic field near the Sun. Hu et al. (2017) and Li et al. (2021) combined the MHD simulation and focused-transport equation to simulate the SEP events observed at 1 au where source particles were injected at the shock according to the 1D DSA solutions. However, these works either mainly focus on the SEP acceleration near the Sun or the transport processes in the IMF. There is still a lack of work studying the spatial and temporal evolution of SEP acceleration from the Sun to 1 au.

In this paper we present results from a 2D kinematic model of the acceleration of SEPs at a CME-driven shock in the IMF. We focus on the geometric effects on SEP acceleration near the Sun and the so-called energetic storm particle (ESP) phase of SEP events. This approach makes no attempt to simulate any particular observed SEP event but focuses instead on fundamental aspects of this problem that provide new insights of the potential geometric effects on SEP acceleration and distribution associated with fast and wide CMEs (e.g., GLEs). We describe the numerical model in the next section with more details provided in the Appendix. The results of simulation are presented in Section 3 and the conclusions are given in Section 4.

## 2. Numerical Model

We investigate SEP acceleration at a 2D spherical shock propagating through the IMF in the solar equatorial plane (the $x$–$y$ plane in Figure 1). We assume a uniform solar wind velocity $V_{sw}$ in the radial direction and the background magnetic field is the Parker spiral magnetic field (Parker 1958):

$$\mathbf{B}(r,\,\theta) = B_0\left(\frac{R_0}{r}\right)^2\left[\hat{r} - \frac{r\Omega_\odot \sin\theta}{V_{sw}}\left\{1 - \left(\frac{R_0}{r}\right)^2\right\}\hat{\phi}\right]. \quad (1)$$

Here $\theta = 90°$, $V_{sw} = 400$ km s$^{-1}$, and $B_0$ is the strength of the field at $r = R_0$, where the field is radial. We take $B_0 = 1.6$ G near the surface of the Sun, which gives a field strength of 5 nT at 1 au. Note that this form of the Parker spiral differs slightly from that in Parker (1958). It assumes that the azimuthal component of the solar wind velocity is $V_\phi = \Omega_\odot R_0^2 \sin\theta / r$, instead of $V_\phi = \Omega_\odot r \sin\theta$ as Parker assumed, and is consistent with the conservation of angular momentum (see Weber & Davis 1967; Priest 1982; Boyd & Sanderson 2003) in the limit of no magnetic stresses. The CME-driven shock (the blue circle in Figure 1) expands outwards from the lower corona to 1 au with a constant speed $V_{sh} = 1500$ km s$^{-1}$, and the shock center is fixed and offset from the center of the Sun at ($x_{sh0}$, $y_{sh0}$) = ($2R_\odot$, 0). This assumption is generally a good approximation for the shocks driven by fast and wide CMEs where the expansion is always dominant in the spatial evolution of the structure (e.g., Kwon et al. 2014).

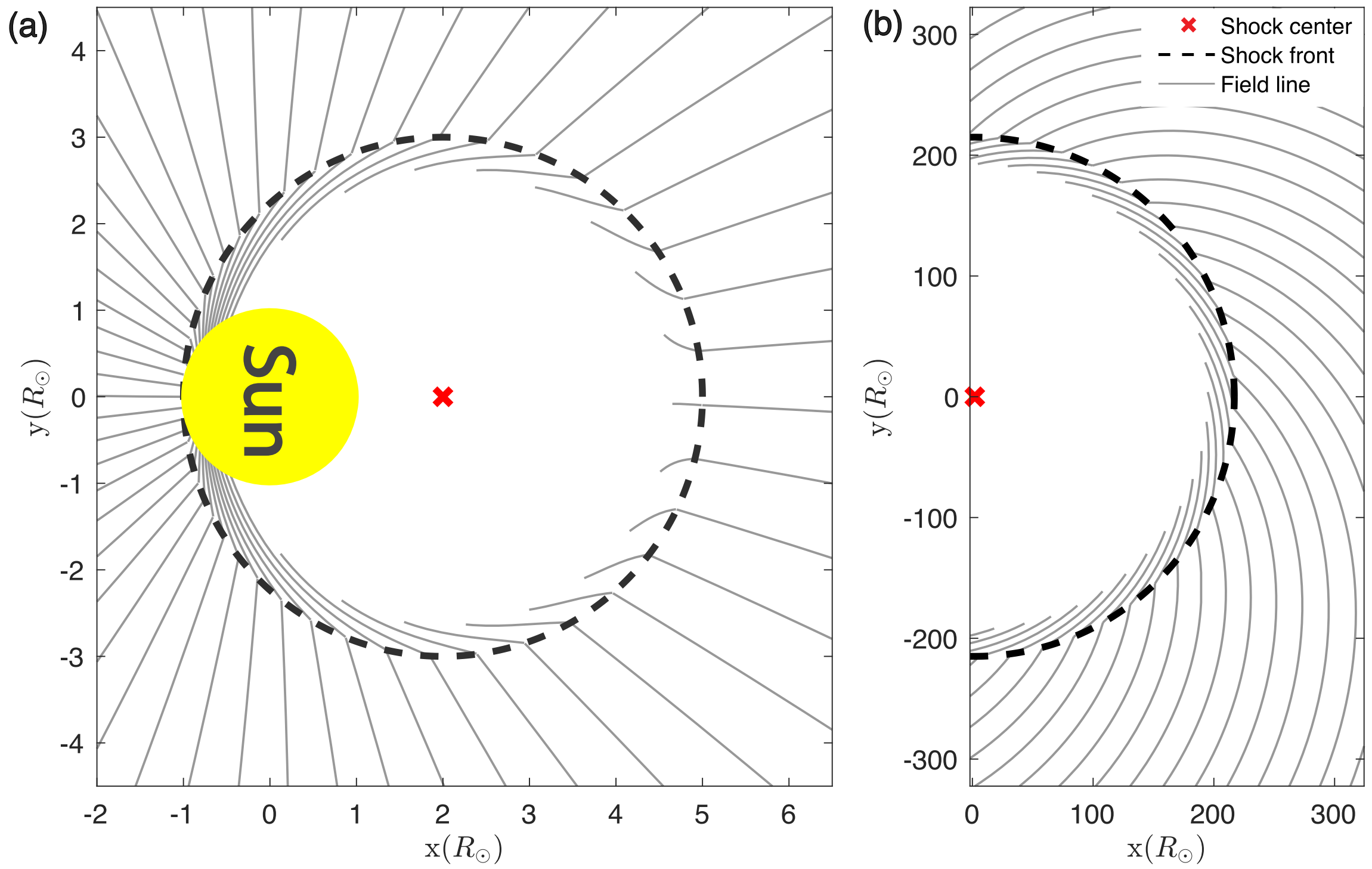

**Figure 2.** The magnetic-field lines in the upstream and downstream of the shock at (a) $R_{sh} = 3\,R_\odot$ and (b) $R_{sh} = 1$ au.

In this fast-mode shock, the transition layer is taken as a planar discontinuity locally since $\Delta_{sh} \sim d_i \ll R_{sh}$, where $\Delta_{sh}$ is the width of the shock and $d_i$ is the ion inertial length. Thus, the macroscopic plasma properties right behind the shock can be obtained from the MHD Rankine–Hugoniot conditions (hereafter jump conditions). We assume that the solar wind plasma expands in a constant velocity in the downstream region with the value determined by the jump conditions, and the magnetic field evolves with the plasma following the ideal MHD equations. Therefore, the plasma properties at any location $L$ ($x$, $y$) in the downstream can be solved in three steps: (1) find the time $t_0$ and location $L_0(x_0, y_0)$ that this infinitesimal volume of plasma was crossed by the shock; (2) apply the jump conditions at $L_0$ to obtain the downstream values; (3) these values evolve from $L_0$ to $L$ through the ideal MHD equations. A detailed description of this method can be found in the Appendix. Figure 2 presents the magnetic-field lines in the upstream and downstream when (a) $R_{sh} = 3\,R_\odot$ and (b) $R_{sh} = 1$ au. Here, we consider the strong shock condition where compression ratio of the shock $s = 4$.

We simulate SEP acceleration by numerically solving the Parker transport equation (Parker 1965):

$$\frac{\partial f}{\partial t} = \frac{\partial}{\partial x_i}\left[\kappa_{ij}\frac{\partial f}{\partial x_j}\right] - U_i\frac{\partial f}{\partial x_i} + \frac{p}{3}\frac{\partial U_i}{\partial x_i}\frac{\partial f}{\partial p} + Q, \tag{2}$$

which describes the evolution of particle distribution function $f$ with the dependence on particle position $x_i$, momentum $p$, and time $t$. $\kappa_{ij}$ is the spatial diffusion tensor; $U_i$ is the bulk plasma velocity and $Q$ is the source. This equation is general enough that contains most of macroscopic transport effects, such as diffusion in the turbulent magnetic field, advection with the background solar wind, and energy change due to the compression or rarefaction of plasma flows. Note that, in our model, the magnetic gradient and curvature drifts are always perpendicular to the $x$–$y$ plane and thus not included here. However, in 3D, these drift could play a role in the acceleration processes depending on the shock geometry and background plasma properties (e.g., Giacalone & Burgess 2010). The only application requirement of Equation (2) is that the pitch-angle distribution of the particles should be nearly isotropic in the plasma frame. Here, we are particularly interested in the distribution and acceleration of SEPs near the shock front or the ESP phase of SEP events where the charged particles have undergone sufficient diffusion and the quasi-isotropic pitch-angle distributions are generally achieved.

The symmetric components of the diffusion tensor $\kappa_{ij}$ can be written in terms of the magnetic-field vector and the diffusion coefficients parallel and perpendicular to the mean magnetic field as (Giacalone & Jokipii 1999)

$$\kappa_{ij} = \kappa_\perp \delta_{ij} - \frac{(\kappa_\perp - \kappa_\parallel)B_i B_j}{B^2}. \tag{3}$$

The antisymmetric components are neglected because of no drift motion in our model. The parallel diffusion coefficient is assumed to scale inversely with the magnetic-field strength $\kappa_\parallel \propto 1/|B|$, which leads to a smaller $\kappa_\parallel$ in the downstream than that in the upstream due to the compression of the field across the shock. The momentum dependence of $\kappa_\parallel$ is obtained from the quasi-linear theory (Jokipii 1971) by assuming the magnetic turbulence in the solar wind following a Kolmogorov power spectrum and given by $\kappa_\parallel \propto p^{4/3}$. The perpendicular diffusion coefficient $\kappa_\perp$ is much smaller than $\kappa_\parallel$ (only several percents) and the ratio $\kappa_\perp/\kappa_\parallel$ is almost independent of particle's energy (Giacalone & Jokipii 1999). Thus, the diffusion coefficients can be written as $\kappa_\parallel = \kappa_{\parallel 0}(B_{1\text{ au}}/B)(p/p_0)^{4/3}$ and $\kappa_\perp/\kappa_\parallel = 0.03$. We consider protons as the source particles of the injection energy $E_0 = 250$ keV and $p_0$ is the corresponding momentum. $B_{1au}$ and $\kappa_{\parallel 0}$ are the averaged magnetic-field strength and the parallel diffusion coefficient of the source particles estimated at 1 au (Giacalone 2015). Note that the diffusion coefficient could be much more sophisticated in the IMF and depends on a number of factors regarding the local turbulence environment and magnetic structures (Zank et al. 1998, 2012; Wang et al. 2022). However, to resolve these contributions, we need to further make the assumptions about the nature of turbulence (e.g., 2D or slab) and the spatial dependence of correlation length that have not been well understood. Here, we choose this relatively simplified form of $\kappa_\parallel$ to avoid these issues and it shows a relatively good agreement with the recent observations from Parker Solar Probe (Li et al. 2022). The source particles are continuously injected to the shock front from the Sun to 1 au and the intensity falls off with the radial distance as $Q \propto r^{-2}$. Generally, the injection efficiency has a strong dependence on the shock-normal angle in the quiet or very weak turbulent magnetic field, known as the "injection problem" (Giacalone & Jokipii 1999; Zank et al. 2006; Guo et al. 2021). However, these limits can be hardly held in most astrophysical environment because of the preexisting turbulence in the upstream, which is believed to play an important role in producing the suprathermal seed population (Giacalone 2005). In this work, we choose the source particles of sufficiently high energy (250 keV) so that the pitch-angle distribution is almost isotropic in the plasma frame and the shock geometry has little impact to the injection rate. Table 1 summarizes the parameters we used in this simulation.

**Table 1**
Simulation Parameters

| | |
|---|---|
| $x_{sh}$ | $2\,R_\odot$ |
| $y_{sh}$ | 0 |
| $R_{\min}$ | $1.5\,R_\odot$ |
| $V_{sh}$ | $1500\text{ km s}^{-1}$ |
| $V_{sw}$ | $400\text{ km s}^{-1}$ |
| $B_0$ | $1.6\,G$ |
| $E_0$ | 250 keV |
| $\kappa_{\parallel 0}$ | $10^{19}\text{ cm}^2\text{ s}^{-1}$ |

The Parker transport equation (Equation (2)) can be written into a Fokker–Planck form and solved by numerically integrating the corresponding stochastic differential equations as (e.g., Kong et al. 2017)

$$\Delta x = r_1\sqrt{2\kappa_\perp \Delta t} + r_3\sqrt{2(\kappa_\parallel - \kappa_\perp)\Delta t}\frac{B_x}{B} + U_x \Delta t + \left(\frac{\partial \kappa_{xx}}{\partial x} + \frac{\partial \kappa_{xy}}{\partial y}\right)\Delta t \quad (4)$$

$$\Delta y = r_2\sqrt{2\kappa_\perp \Delta t} + r_3\sqrt{2(\kappa_\parallel - \kappa_\perp)\Delta t}\frac{B_y}{B} + U_y \Delta t + \left(\frac{\partial \kappa_{yy}}{\partial y} + \frac{\partial \kappa_{xy}}{\partial x}\right)\Delta t \quad (5)$$

$$\Delta p = -\frac{p}{3}\left(\frac{\partial U_x}{\partial x} + \frac{\partial U_y}{\partial y}\right)\Delta t \quad (6)$$

where $r_1$, $r_2$, and $r_3$ are different sets of normalized random numbers that satisfy $\langle r_i \rangle = 0$ and $\langle r_i^2 \rangle = 1$. The trajectories of pseudo-particles are resolved by numerically integrating the stochastic differential equations. Due to the strong velocity gradients inside the transition layer, pseudo-particles are accelerated at each time they cross the shock. The spatial size of the derivative is $10^{-5}\,R_\odot$ and the time step we used here is 0.01 s, which is small enough to ensure DSA is working properly. The outer boundary of the model is 2 au from the origin (center of the Sun) and the inner boundary is determined by the iteration size in the Appendix, which is about 15%–20% of the shock radius in the downstream (see Figure 2). We calculate the return probability as the pseudo-particles pass the upstream or downstream boundaries to determine if they should be removed from the simulation (Jones & Ellison 1991). We also use a particle splitting technique (Giacalone 2005) to improve the statistics.

## 3. Results

Figure 3 presents the spatial distribution of SEP number density as the shock propagates through the Parker spiral magnetic field at different shock radii, i.e., 3, 10, 30, 100 $R_\odot$, and 1 au, respectively. We find that the energetic particle density varies significantly in both space and time, which is closely related to the local shock geometry ($\theta_{B_n}$) and radial distance. For example, at early times, the "hot regions" of high-energy SEPs form along the flanks of the shock and very few particles are efficiently accelerated at the shock nose (Figure 3(a)). This is because the shock is quasi-perpendicular near the two flanks and nearly parallel at the nose. The rate of acceleration is highest at the quasi-perpendicular shock and lowest at the nose, and, thus, for a given simulation time, this leads to these "hot regions" at the flanks and few particles at the nose. Because the magnetic field near the Sun is close to radial, this leads to a symmetric shock geometry about $y = 0$, and thus the similar particle distributions along the two flanks (e.g., Figure 3(a)). However, as the shock propagates outward, the connection between the shock front and Parker spiral magnetic field causes $\theta_{B_n}$ to vary in time.

For the convenience of discussion, we define the shock nose as the intersection between the shock front and $y = 0$ ($\phi = 0°$), and the west and east flanks as the parts of the shock at $y > 0$ and $y < 0$ respectively (see Figure 3(a)). In Figure 3(b) and (c), we find that the west flank gradually becomes a preferred region for accelerating SEPs, which can be explained by considering the shock geometry in the spiral magnetic field. Generally, the west flank has a larger averaged $\theta_{B_n}$ than that in the east flank, corresponding to a higher acceleration rate. This effect will be eliminated as the shock radius increases in which case $\theta_{B_n}$ becomes almost a constant along the shock front (e.g., Figures 3(d) and (e)) and finally we obtain a relatively uniform SEP distribution at 1 au.

We now consider the longitudinal dependence of the particle energy spectra at different shock radii (and time). The top three panels of Figure 4 present the number density of SEPs detected near the shock at $\phi = W60° \pm 2°$ (upper most plot), $E60° \pm 2°$ (middle plot at the top of this figure), and $0° \pm 2°$ (bottom of the three plots at the top) as a function of particle's energy and the time after injection ($\Delta t = t - t_{\rm inj}$), where $t$ is the current time step and $t_{\rm inj}$ is the initial time when the particle was released to the shock. $\Delta t \sim 0$ indicates that the particles were just added to the system while $\Delta t > 0$ means that the particles were injected earlier and traveled along with the shock to the current location. A "V-shape" distribution can be identified in the top three panels of Figures 4(b)–(d) as the shock moves outwards. We find that most of high-energy particles are injected relatively early (large $\Delta t$) as they have more chance to interact with the shock and achieve continuous acceleration from the injected site to current location. The newly released particles (small $\Delta t$) are generally of lower energy because of the limited acceleration time. The maximum energy of these "fresh particles" (say $\Delta t < 1$ hr) decreases significantly with radial distance from more than $10^3$ MeV at 10 $R_\odot$ (Figure 4(a)) to around the injection energy 250 keV at 1 au (Figure 4(d)). This trend reflects the local acceleration rates at different shock radii: very rapid acceleration near the Sun but almost no acceleration at 1 au. We also include adiabatic cooling in our model due to the expansion of the solar wind. This is always present during the acceleration process, even though the acceleration rate is considerably higher than the cooling rate, especially near the Sun. The particles that are advected into the downstream and no longer come back to the upstream will keep cooling down in the expanding solar wind, and thus form the other half of "V-shape" distribution below the injection energy.

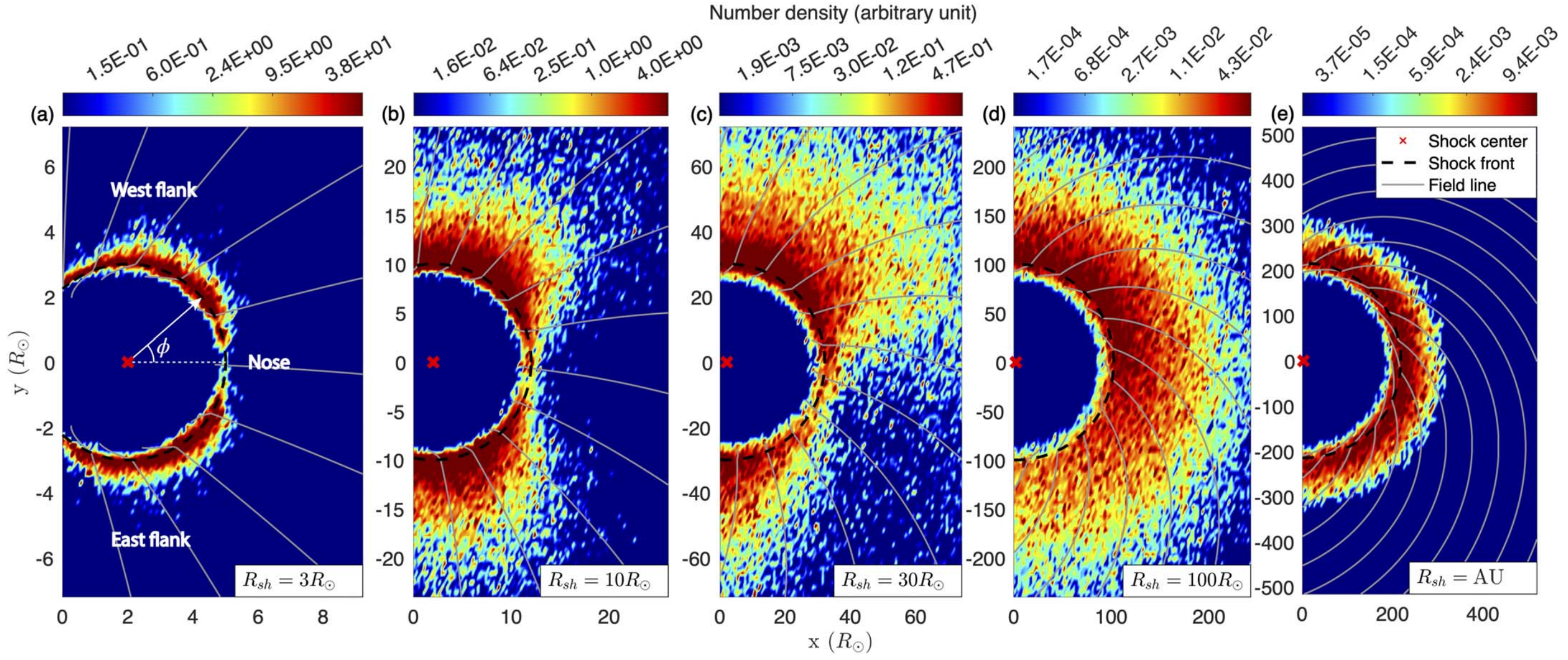

**Figure 3.** The spatial distribution of SEPs with energy $E > 10$ MeV as the shock propagates in the Parker spiral magnetic field (gray lines) at different shock radius $R_{sh}$: (a) 3 $R_\odot$, (b) 10 $R_\odot$, (c) 30 $R_\odot$, (d) 100 $R_\odot$, and (e) 1 au. The red "x" indicates the shock center, fixed at $x_{sh} = 2\ R_\odot$ and the black dashed circle presents the position of shock front.

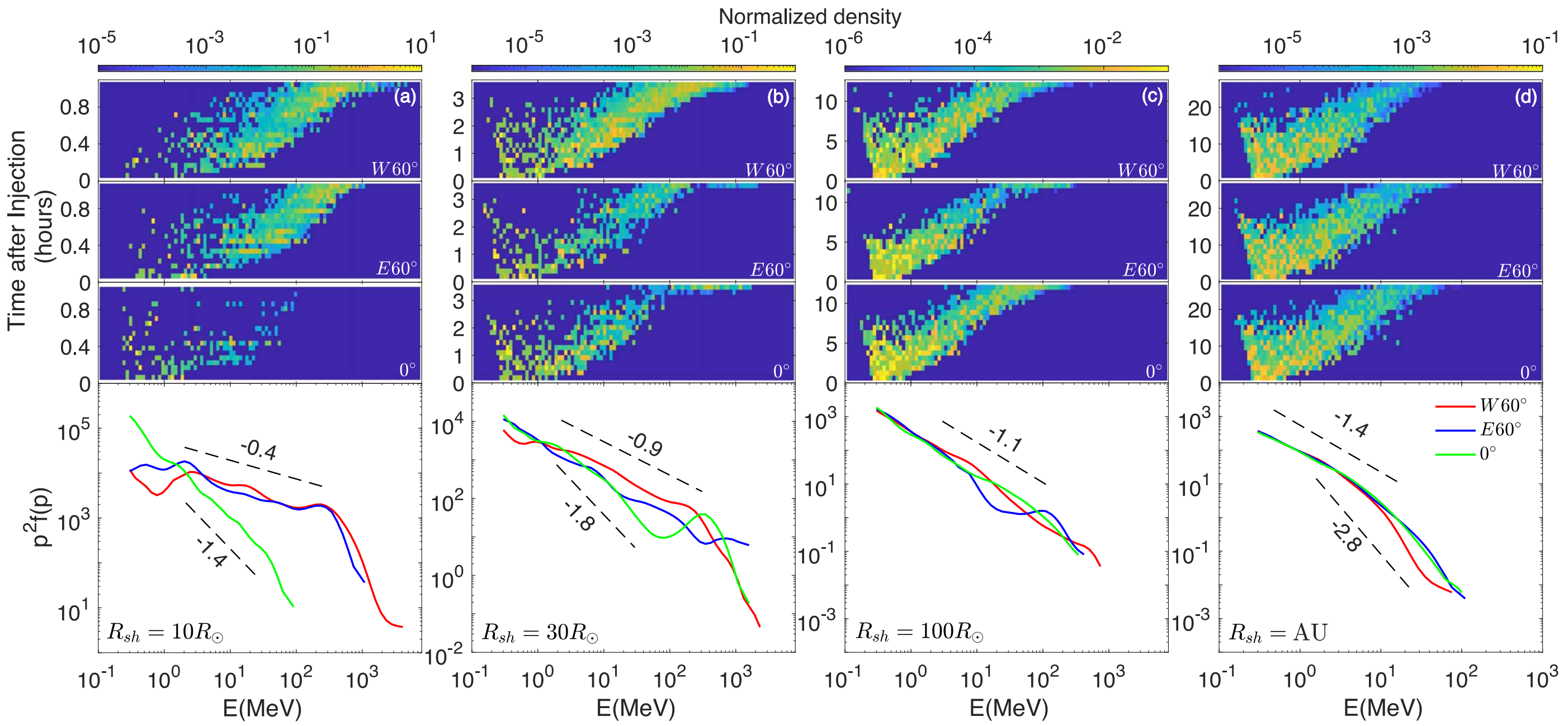

**Figure 4.** Energy spectra at the west ($W60° \pm 2°$) and east ($E60° \pm 2°$) flanks of the shock as well as shock nose ($0° \pm 2°$) when the shock radius reaches (a) 10 $R_\odot$, (b) 30 $R_\odot$, (c) 100 $R_\odot$, and (d) 1 au. The top three panels present the number density of SEPs detected at different locations along the shock front as a function of particle's energy and the time after injection. The button panel shows the corresponding energy spectra of the particles in the top three panels. The dashed lines provide a reference of the flattest and steepest spectra among a certain energy range.

The bottom panels in Figure 4 show the energy spectra of the corresponding particles in the top three panels. The dashed lines provide references of the spectral indices. Note these spectra at the flanks in Figure 4(a) are even harder than the spectrum at a strong shock ($s = 4$) predicted by DSA. This is because in the DSA prediction the particle spectrum is calculated right behind a planar shock in the downstream which will give a "-1" spectral index with the strong shock limit. However, the spectra in our work are obtained from both the up- and downstream particles among a small angular range ($\Delta\phi = \pm 2°$) so that the earlier accelerated particles in the upstream will also contribute to the spectra. This approach can help us understand how the ESPs evolve in space and time (as shown in the top three panels of Figure 4) but also gives rise to a flatter spectrum near the Sun. In Figure 4, we can find that the maximum energy of the spectrum decreases as the shock moves outward. This is mainly caused by two reasons: (1) the large diffusion coefficients away from the Sun make the shock even harder to accelerate SEPs to very high energies; (2) the highest-energy particles generated near the Sun can escape the trap of the shock and travel across the boundaries very shortly, which leads to this population gradually reducing in time. As a result, the spectral indices and cutoff energy generally decrease with the radial distance (Figures 4(a)–(d)). The spectra at shock flanks are much harder than that at the nose in the region near the Sun but gradually converge as the shock expands outward, which is consistent with the shock geometry evolution discussed above. Both the double power law (e.g., west and east flanks in Figure 4(a)) and the dip features (e.g., nose in Figure 4(b)) can be found in the simulation. With considering the top three panels, we notice that both features are related to the evolution of shock geometry and contributed by the earlier injected particles as well as the locally accelerated components. For example, if we compare the third and fourth panels of Figure 4(b) we can find that the low-energy part of the spectrum (<70 MeV) at the shock nose (green line) is mainly produced by the local quasi-parallel shock, which leads to a relatively steep spectrum. However, if we trace this field line back to the initial position of the shock ($t = 0$), the intersection between the field line and shock front will shift $\sim 5°$ counterclockwise toward the west flank. This means that the high-energy part of the spectrum (>70 MeV) is generated by the SEPs from a preferential acceleration region (west flank) near the Sun, and thus give rise to a dip–bump structure. In other words, the in situ measurements of SEPs may not only related to the local shock and solar wind properties but also dominated by the plasma environment in the early state near the Sun. These results reveal that the evolution of SEPs as the shock propagates from the Sun to the Earth is crucial for the studies and prediction of large SEP events, and both acceleration and transport processes should be included in the consideration.

To better understand the SEP distribution at 1 au, we investigate the initial conditions of these particles when (and where) they were injected. Figure 5 presents the injection location of the particles that were accelerated to high energies ($E > 10$ MeV) and then observed at 1 au. The white "X" sign indicates the location of shock center. We see that most SEPs (that later become high-energy particles at 1 au) are injected at the west flank near the Sun. This confirms that west flanks is the preferred region to accelerate high-energy particles due to the relatively large $\theta_{B_n}$. As the shock propagates outward, the distribution of the source particles gradually spreads over a wider range along the shock front and hardly any particles can be found beyond 100 $R_\odot$. This is because (1) the source intensity decreases with radial distance ($Q \propto r^{-2}$) and (2) the diffusion coefficients increase and become relatively uniform far from the Sun, leading to a lower acceleration rate. The SEP distribution at 1 au is a result of contributions from both of these factors.

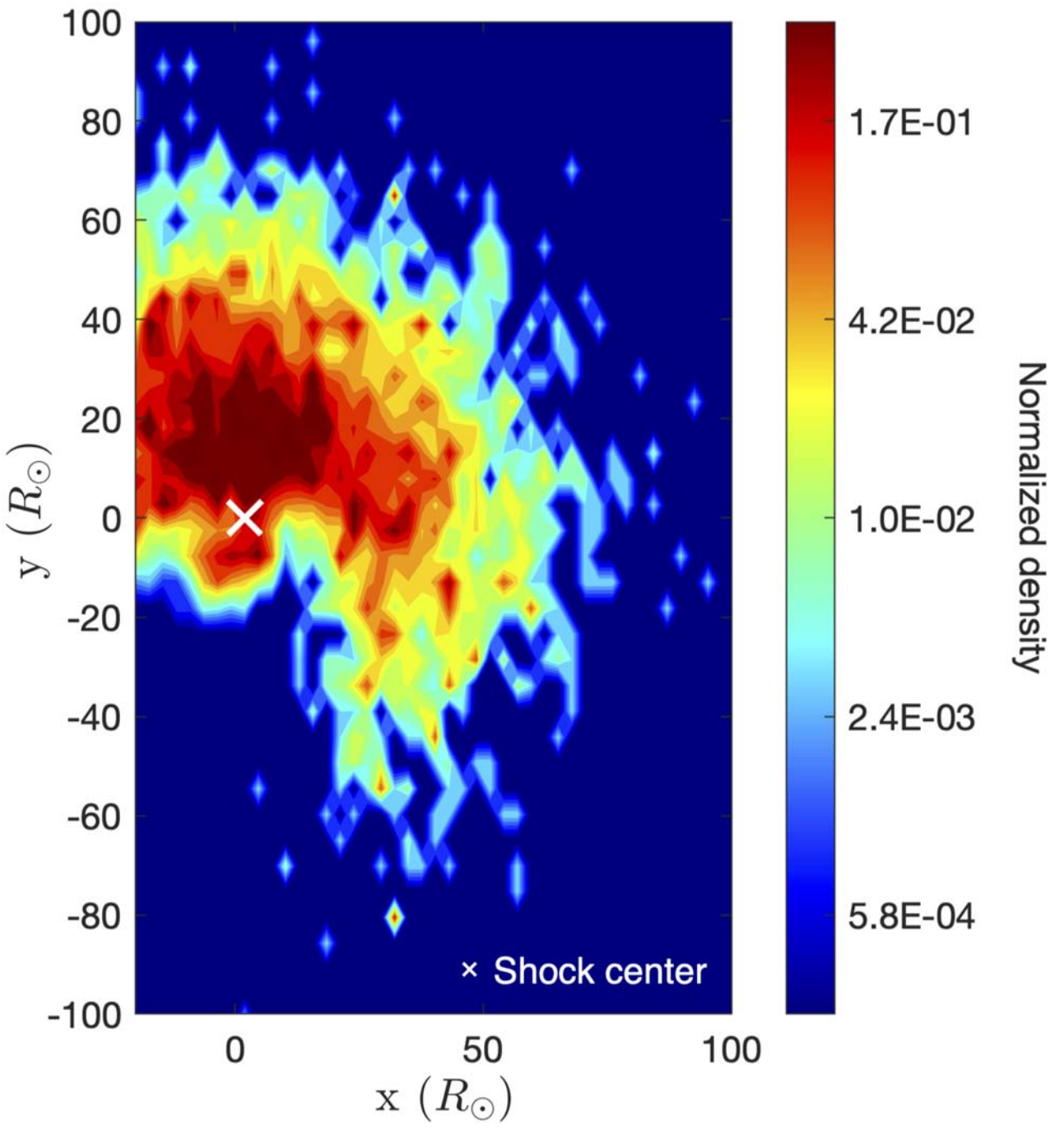

**Figure 5.** The injection position of high-energy particles (>10 MeV) observed at 1 au. The white "X" denotes the location of shock center.

## 4. Conclusions

In this study, we developed a 2D kinematic model to study the SEP acceleration at a spherical shock with the shock geometry evolving in both space and time as it propagates in the interplanetary space from a few solar radii to 1 au. The shock expands outwards from its center with a constant speed, and the center is fixed in time and offset from the center of the Sun. The assumed source of particles was monoenergetic (taken to be 250 keV), distributed uniformly along the shock, with a flux across the shock that falls off as $1/r^2$, where $r$ is measured relative to the center of the Sun. Our model was a numerical solution to the the Parker transport equation, which provides the distribution of particles accelerated at the shock and transported away from it. We are particularly focusing on the distribution of high-energy particles in the shock region from a distance near the Sun and close to 1 au (ESPs) as the shock propagates through the IMF.

The primary conclusions from our study are

1. The local shock/magnetic-field morphology plays an important role in the resulting distribution of SEPs and their acceleration at the shock.
   (a) Most high-energy SEPs are accelerated close to the shock flanks instead of the shock "nose" (along the Sun–Earth line) in the region near the Sun.
   (b) The west flank of the shock is where there is a preferential acceleration of the particles to high energies compared to the east flank and shock "nose" due to a larger $\theta_{B_n}$ in this region, and a higher acceleration rate of the particles.
2. >10 MeV SEPs are accelerated very rapidly at the shock near the Sun but hardly any high-energy particles are accelerated locally at the shock when it crosses 1 au. This is caused by the variation of the diffusion coefficient, being much smaller near the Sun and much larger far from the Sun, and the fact that the acceleration rate depends inversely on the diffusion coefficient.
3. The simulated energy spectra vary significantly along the shock front at different shock radii (at different times in the evolution of the shock). Moreover, both double power law and dip features are seen in the simulated energy spectra.

These results emphasize that in large SEP events, the particle intensities are not only determined by the local shock structures and plasma environment but also strongly dependent on these properties near the Sun. Both acceleration and transport processes should be considered to study the spatial and temporal evolution of SEPs in the IMF. We hope this new model will provide some insights into wide spread SEP events and help us a better understanding of the multispacecraft observation, especially those close to the Sun, such as Parker Solar Probe and Solar Orbiter.

We are grateful for helpful discussions with Jozsef Kota and Federico Fraschetti. The project was supported in part by NASA under grants 80NSSC20K1815 and 80HQTR21T0005. J.G. acknowledges support from the IS⊙IS instrument suite on NASAs Parker Solar Probe Mission, contract NNN06AA01C. F.G. also acknowledges support in part by NASA grant 80HQTR21T0087 and 80HQTR21T0117.

## Appendix
## Flow Velocity and Magnetic Field in the Downstream

As noted in Section 2, we assume the magnetic field is frozen into the plasma and the solar wind flow in the downstream region of the shock remains in a steady state (relative to the shock) after it starts to be compressed by the shock. With these assumptions, we can obtain $\vec{U}$ and $\vec{B}$ at any point in the downstream region based on the jump conditions at the time and location where the shock passed this point of plasma, and how the flow expands in the downstream.

First, we apply the jump conditions to calculate the flow velocity and magnetic field downstream of the shock based on those in the upstream. A schematic of the jump conditions is shown in Figure 6(a) where the subscripts 1 and 2 indicate the properties in the upstream and downstream respectively. Both $\vec{U}$ and $\vec{B}$ are considered in the local shock frame. Here, $\vec{U_1}$ is a combination of shock and solar wind velocities ($\overrightarrow{U_1} = \overrightarrow{V_{sh}} - \hat{n} \cdot \overrightarrow{V_{sw}}$), and $\vec{B_1}$ is the nominal Parker spiral magnetic field (Equation (1)). We resolve $\vec{U_1}$ and $\vec{B_1}$ into the direction parallel and perpendicular to the shock normal $\hat{n}$ ($U_{1\parallel} = U_1 \cos\delta_1$, $U_{1\perp} = U_1 \sin\delta_1$, $B_{1\parallel} = B_1\cos\theta_1$ and $B_{1\perp} = B_1 \sin\theta_1$), and the downstream properties are given by the jump conditions as (Decker 1988)

$$U_{2\parallel} = U_{1\parallel}/s \tag{A1}$$

$$U_{2\perp} = U_{1\perp} + \frac{(s-1)\cos\theta_1 \sin\theta_1}{M_{A1}^2 \cos^2\delta_1 - s \cdot \cos^2\theta_1} U_{1\parallel} \tag{A2}$$

$$B_{2\parallel} = B_{1\parallel} \tag{A3}$$

$$B_{2\perp} = \frac{M_{A1}^2 \cos^2\delta_1 - \cos^2\theta_1}{M_{A1}^2 \cos^2\delta_1 - s \cdot \cos^2\theta_1} \cdot s \cdot B_{1\perp} \tag{A4}$$

where $s$ is the plasma compression ratio across the shock and $M_{A1}$ is the Alfvénic Mach number in the upstream region.

To obtain $\vec{B}$ and $\vec{U}$ in the downstream, we use an iterative method to solve the time and position when the shock crossed a plasma point of interest. Here, we consider a spherical coordinate system of the origin fixed at the shock center (Figure 6(b)). Given any point $L(t)$ in the downstream of the shock, the shock crossing position $L_0(t_1)$ can be solved if we assume the flow velocity only has the radial component $U_{2\parallel}$:

$$L_0(t_1) = R_{sh}(t_1) = \frac{R_L(t)V_{sh} - R_{sh}(t)U_{2\parallel}}{V_{sh} - U_{2\parallel}} \tag{A5}$$

where $R_L(t)$ is the radial distance of point $L(t)$ and $U_{2\parallel}$ is the radial velocity obtained from the jump conditions applied at $L(t)$. $L_0(t_1)$ is an initially guessed solution. Then, we solve $\vec{U}_2(t_1)$ at $L_0(t_1)$ through the jump conditions and convert this guessed plasma point back to current time step $t_1 + \Delta t$: $\vec{L}(t_1 + \Delta t) = \vec{L}_0(t_1) + \vec{U}_2(t_1)\Delta t$, where $\Delta t = (R_{sh}(t) - R_{sh}(t_1))/V_{sh}$. By comparing the spatial difference between $\vec{L}(t_1 + \Delta t)$ and $\vec{L}(t)$, we can obtain the correction term $\Delta\vec{l}$ and apply it to $\vec{L}_0(t_1)$ to get another guessed-point $\vec{L}_0(t_2)$ of improved accuracy. If we apply the jump conditions at $\vec{L}_0(t_2)$ and repeat this routine $N$ times, we can finally solve the position $\vec{L}_0(t_N)$ ($N \to \infty$) where the shock just passed by this particular point of plasma of interest. Note that the efficiency of the iteration can be significantly improved by applying a reasonable accuracy. For example, in this study we consider the acceptance $\Delta\vec{l} < 0.01 V_{sh}\Delta t$ and the iterative times are generally less than 10.

After perturbed by the shock, the magnetic field will evolve with the expanding plasma following $B_\parallel \propto 1/r^2$ and $B_\perp \propto 1/r$. Here we only consider the radial expansion of the plasma flow in the downstream since $U_{2\parallel} \gg U_{2\perp}$. So that $\vec{B}$ at $\vec{L}(t)$ can be solved as

$$B_{L\parallel} = B_{L_0\parallel}\left(\frac{R_{sh}(t_N)}{R_L(t)}\right)^2 \tag{A6}$$

$$B_{L\perp} = B_{L_0\perp}\frac{R_{sh}(t_N)}{R_L(t)} \tag{A7}$$

where $B_{L_0\parallel}$ and $B_{L_0\perp}$ are the downstream magnetic field at $\vec{L}_0(t_N)$. Figure 2 presents the geometry of the IMF lines in the both up- and downstream of the shock when the shock is near the Sun and has a radius of 1 au.

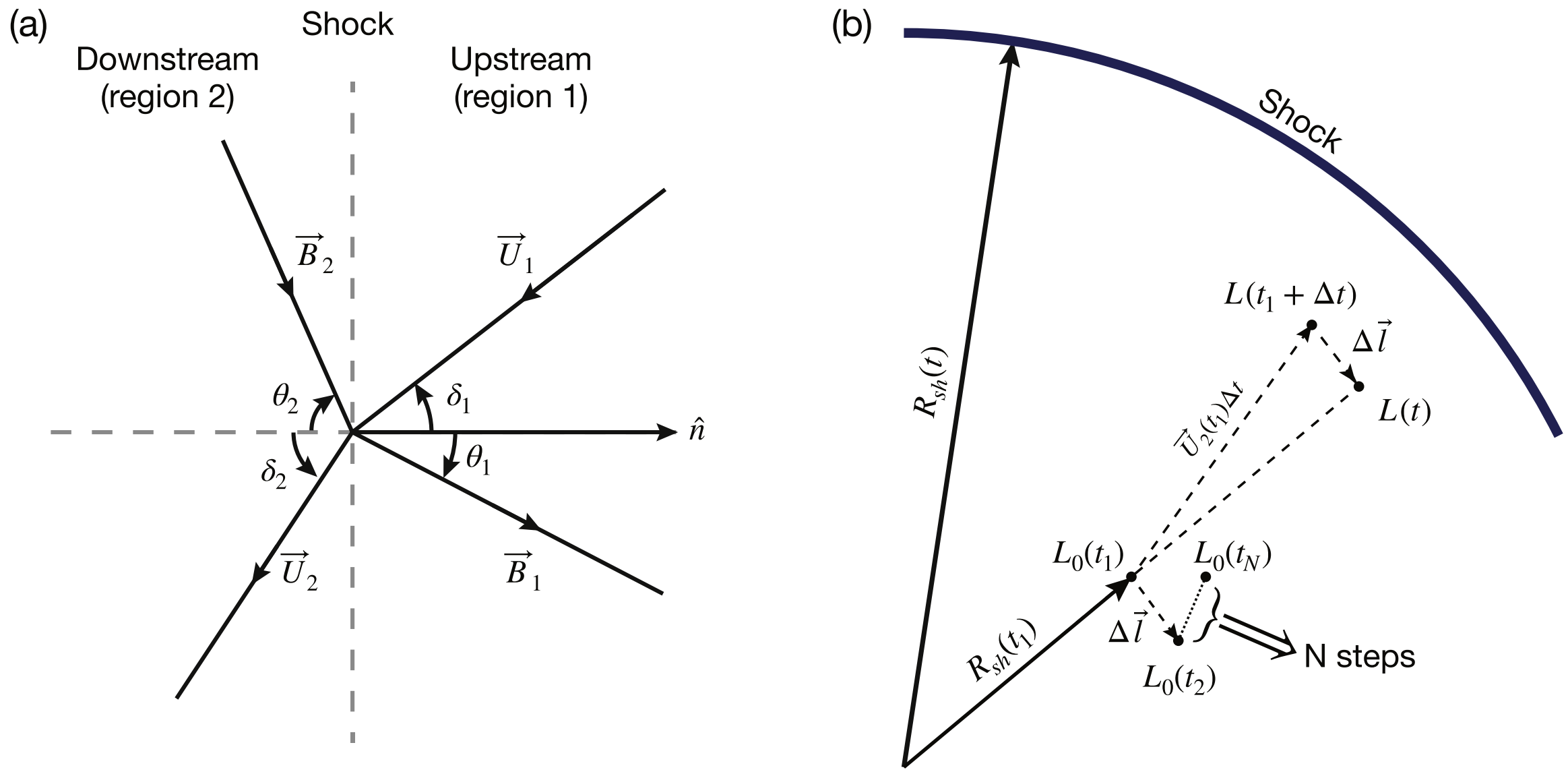

**Figure 6.** The schematics of (a) the MHD Rankine–Hugoniot conditions and (b) the geometry used for solving the downstream magnetic field.

## ORCID iDs

Xiaohang Chen https://orcid.org/0000-0003-2865-1772
Joe Giacalone https://orcid.org/0000-0002-0850-4233
Fan Guo https://orcid.org/0000-0003-4315-3755